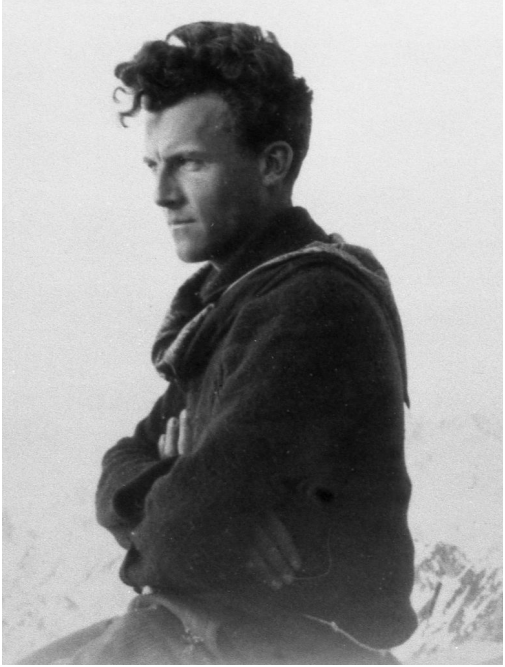

*"The very struggle towards the summits is enough to fill a man's life."*

Albert Camus

# Joseph Devaux (1902-1936)
## Meteorologist at Pic du Midi Observatory

Emmanuel Davoust[1] and Jean-Paul Meyer[2]

**Abstract.** Joseph Devaux, who held a position of assistant-meteorologist at Pic du Midi Observatory, had an outstanding career. He devoted his life to the Pic and to scientific research. A pioneer of the study of snow and glaciers, he also pursued research in many fields of atmospheric physics. He was the physicist with Commandant Charcot's polar expeditions but was lost at sea (when the *Pourquoi-Pas?* sank) before he was able to use his talents to the full.

**Résumé.** Joseph Devaux, qui était assistant-météorologiste à l'Observatoire du Pic du Midi, a eu une carrière remarquable. Il a consacré sa vie au Pic et à la recherche scientifique. Pionnier de l'étude de la neige et des glaciers, il a aussi poursuivi des recherches dans de nombreux domaines de la physique de l'atmosphère. Il était physicien dans les expéditions polaires du Commandant Charcot, mais a disparu en mer (lorsque le *Pourquoi Pas?* a sombré) avant d'avoir pu donner toute sa mesure.

**Birth of a Vocation**

Joseph Devaux was 18 years old when he discovered the Pic du Midi in August 1920. He was accompanying his father who had, for the previous several years, been regularly coming there to study the plants at the summit with Joseph Bouget, the observatory's botanist. He learned about the observatory and its inhabitants, including Sylvain Latreille, its meteorologist since 1889, and Marcel Dauzère, the director's nephew. Being about the same age, they went together on excursions in the area, one day getting lost in the fog. Joseph Devaux amused himself by throwing parachutes from the summit in order to study winds. This was his first scientific experiment. He descended from the Pic full of enthusiasm, wishing to live and work up there. The rest of his life would be dedicated to the Pic.

---

1   Observatoire Midi-Pyrénées, Toulouse, France
2   Météo-Fance, Centre national de recherches météorologiques, Toulouse, France

He studied at the Faculty of Science in Bordeaux. When it was time to think about his military service, he tried to become a military meteorologist at the Pic du Midi, as others had done before him. But the army's meteorological service had no openings and could not accept him.

In August and September of 1924, a job as a watchman-guide at the Pic du Midi provided him with an opportunity to better know the observatory, its inhabitants, life at the summit, as well as to admire the range of the Pyrenees, its sunsets and nights. Besides guided tours, he had to carry out some routine meteorological observations. While waiting for the midnight duty, he dreamt of being a vulture flying effortlessly over the Pyrenees. "Wings extended, feathers spread out, beak to the wind, I glide calmly and regularly under the contrary puffs of the wind that raises me up." During his free time, he went on hikes, but also took up again his wind studies with parachutes.

Joseph Devaux did his military service as a meteorologist at Saint-Cyr, Cazeaux, Chartres and Mainz (Germany) from the autumn of 1924 to 1926.

When one of the two meteorological positions had an opening in December 1926, he obtained the job temporarily. He spent Christmas at the summit, alone with a young student. His solitary contemplation of this exceptional night in this exceptional framework inspired him to ask several beautiful questions:

> "Who will describe these overhanging crests of rounded, friable snows that lead to ridges of black and red rock which crumble as soon as they are touched?
>
> Who will talk of the plunging gorges where snow accumulates indefinitely and where one sinks chest-deep into depthless snow?
>
> Who will climb these hardened, icy slopes where an uncertain foot vainly tries to hold onto this surface as slippery as glass?
>
> Who will feel these motionless, opaque fogs amidst which one walks indefinitely without distinguishing the ground from space and with no hope of ever reaching a goal?"

Devaux once again left the Pic du Midi on February 1$^{st,}$ 1927 to prepare a graduate degree at the Faculty of Science in Bordeaux. He returned as watchman-guide in July 1927. At that time, he already had a research project – studying air currents and snow around the Pic du Midi, and he also had a bursary from the *Caisse Nationale des Sciences*. He was permanently hired as an assistant meteorologist at the observatory on January 1$^{st}$, 1928.

His principal job was to participate in the regular meteorological measurements and to ensure their continuity. Every three hours, and in all kinds of weather, it was necessary to go out to the "blockhaus", read the thermometers, barometers, the hygrometer and the pluviometer. Started in 1873, these observations are an exceptional data base about evolution of the climate, temperature and ozone level.

During his time off, with his scientific mind permanently alert, he could analyze the natural phenomena he observed from the Pic. He was the first permanent resident of the Pic du Midi to do science outside routine observations. In 1930 he was joined by Hubert Garrigue, like Devaux a trained physicist, with whom he would form a close-knit team, underpinned by a deep friendship.

Joseph Devaux devoted his main work to studying snow. He was one of the first scientists, if not the first, to measure and study the physical characteristics of névés and glaciers, principally those of the Pyrenees, but also those of the Alps and Greenland.

To do this, he had to adapt or build himself various measuring devices: thermometers, photometers, actinometers, etc. "I am making myself a spectrograph. It is faster and cheaper than to order one", he said one day to a journalist who asked him what he was doing.
He first of all measured the density, then the temperature, of a layer of snow at varying depths. Finally, he measured the thermal conductivity of the snow and the absorption of solar radiation by layers of snow and ice. This would be the basis of the thesis he presented at the Sorbonne on April 24, 1933: "Radio-thermal economy of fields of snow and glaciers".

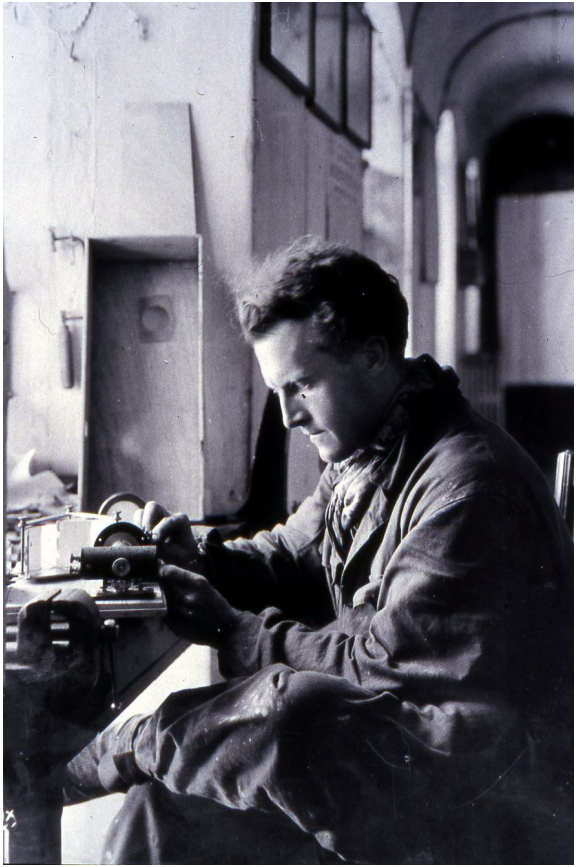 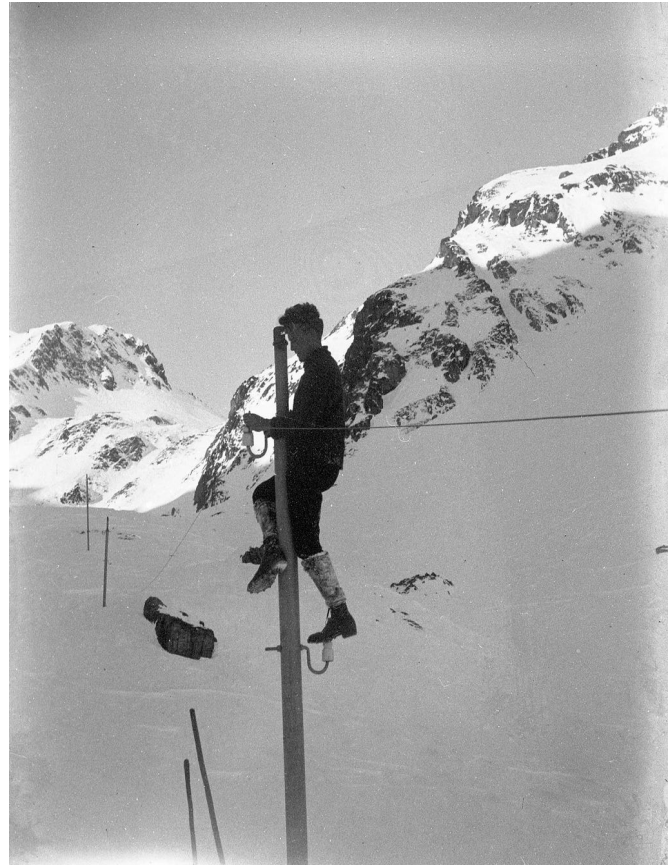

*Joseph Devaux : working at Pic du Midi Observatory, fixing the telephone line to the Observatory.*

His observations allowed him to quantify snow's densification, leading to the formation of névés and glaciers, as well as proposing an empirical formula linking the thermal conductivity of snow and its density. His photometric measurements showed the great variability in the albedo of snowy or icy surfaces, in other words, according to their age, as well as the exponential decrease of solar radiation penetrating snow as a function of its depth.

Still today, Devaux's results are the basis of equations used in modelling snow cover, whether to forecast the risk of avalanches or to represent snow cover in forecast and climate models.

Joseph Devaux was also interested in atmospheric radiation. He perfected a spectrograph using a resonance electrical set-up and automating its operation. Hence, he was among the first to observe the infrared spectrum of the atmosphere. In 1931, he wrote: "We therefore have strong reason to believe that the band towards 10 microns that I found in the sky spectrum is an emission band of atmospheric ozone." He managed to deduce from these measurements the average temperature of ozone. However this temperature is very low and does not allow very precise measurements.

He managed to photograph luminous columns and provided an explanation of this atmospheric optical phenomenon. He often observed the green ray at sunset and gave a precise scientific explanation.

Joseph Devaux participated in his colleagues' observations and experiments. He carried out the photometric and actinometric study of a lunar eclipse with František Link, a Czech astronomer who spent long periods at the Pic in the 1930s. He helped Link and Marcel Hugon, another regular visitor of the Pic, to take direct measurements of atmospheric absorption, from one peak of the Pyrenees to another. With Pierre Idrac, particularly famous for his role in the invention of radio probes, he made fast-motion films of the clouds surrounding the Pic du Midi in 1933. Finally, as a result of his first experiments with parachutes and his fascination with birds of prey, he showed the existence of vertical ascending air currents around the Pic du Midi and their use by vultures.

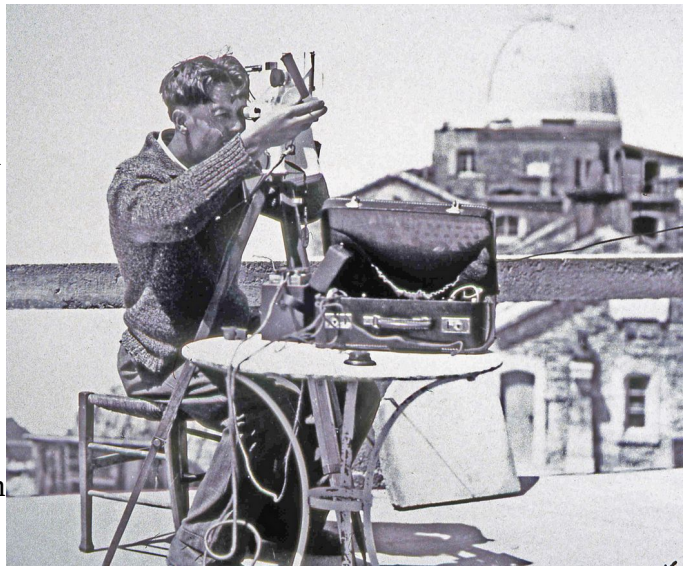

*František Link at Pic du Midi Observatory*

Climate conditions at the Pic du Midi are very similar to those at the poles and, like several other scientists from the Pic before and after him, Devaux participated in polar scientific expeditions. He went three times to Greenland aboard the *Pourquoi-Pas?* with Commander Charcot in the months of July, August and September 1932, 1933 and 1936. With Professor Charles Maurain, he took measurements of various parameters of the atmosphere. He also continued his studies of ice, particularly the icebergs and glaciers around Scoresbysund. He also took advantage of his mountain-climbing skills to participate in exploration and hence contribute to mapping then still unknown regions such as Liverpool Land, north of the Arctic Circle.

The third expedition came to a tragic end off the Icelandic coast on September 16, 1936. Joseph Devaux disappeared, aged 34, during the sinking of the ship of which a single crewmember survived.

**A Difficult Daily Life**

 It is certain that the exceptional position of the Pic du Midi determined Joseph Devaux's vocation and played an important role in his daily life and that of his companions. "But our professional concerns take over, and all those beauties end up translated as optical numbers and calculations", he confided one day, "without however leaving us insensitive." We should add that the Pic's location does not have only advantages. When the weather is bad, its inhabitants are in the clouds, often in storms, and in any case with the impossibility of getting out. In the long term this can become trying. Furthermore, the altitude makes sleeping difficult.

It should also be said that the Pic in 1930 was not the comfortable, easily-accessed site it is today. Hubert Garrigue used to say that "the Pic is a ten-day walk from France", a walk that was very difficult in winter, that had to end by a climb up the Black Rocks holding onto a metal cable. Supplies were brought mostly during the two months of summer, on muleback. The rest of the year, fresh food supplies had to be carried up by men when weather permitted. "One winter we ate canned food for six weeks without a change. We began to worry about getting scurvy", laughed

Joseph Devaux. The cablecar arrived much later, only in 1952, completely transforming life at the Pic.

The buildings, which had not been maintained during the Great War, were completely restored at the instigation of the director Camille Dauzère (Davoust et Damiens, 1995). Central heating and a large electric generator were installed and a new laboratory building installed against the "blockhaus" on the terrace. But the inhabitants' living space remained limited. They lived in a single building. Its walls were solid and a meter thick but it was badly-ventilated. The windowless rooms gave onto the corridor where work was done in the wide openings of the casement windows. The Baillaud Dome, with its 50 centimeter telescope, was used only in summer. Permanent staff was limited to two meteorologists and two employees, sometimes only one if the other was off duty. The site was too high to pick up radio programs and the only leisure activity was reading. Mail was brought once a week (if weather conditions permitted!) and was the inhabitants' principal connection with the outside world.

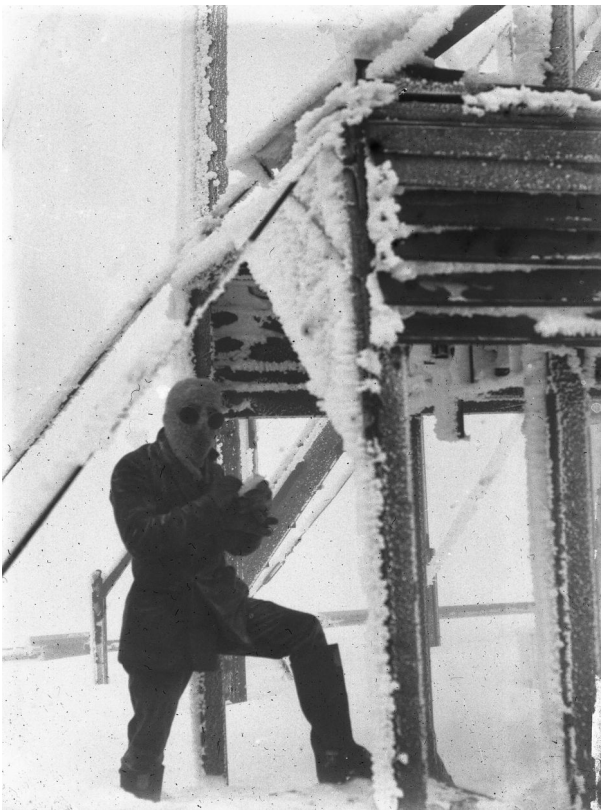
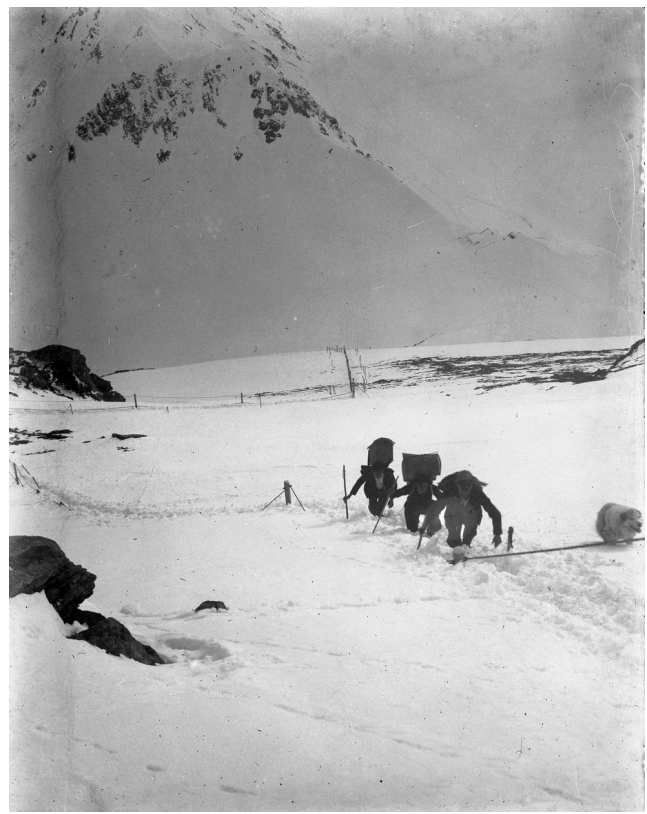

*Reading the meteorological measurements*  *The porters climbing the slope of the Pic du Midi*

These particular living conditions had their effect on scientific work, provoking surprise and admiration from Charles Fabry about Joseph Devaux's work. "It is remarkable that he can carry out such delicate studies with such limited means and with nobody's help." As Hubert Garrigue noted in his diary in July 1936, this austere life also surprised tourists who did not understand it. "They live here alone all year long?" In response, Devaux silently shrugged his shoulders.

It takes much strength of character to live in such conditions, to face storms, isolation, the lack of oxygen. Despite this difficult and solitary daily life, Devaux showed himself to be a warm companion. A man of feeling, he inspired friendship in those who met him, both at the Pic or elsewhere. The zoologist Pierre Drach wrote about him: "I will tell you who was the well-loved companion of many mountain expeditions in Greenland and Iceland. Above all he loved the solitude of his observatory at the Pic du Midi that took him away from the sad mediocrities of daily life. Everything about Devaux was marked with distinction and a rare quality. His natural modesty and his simplicity prevented no-one from seeing in him an elite spirit and a superior man."

**The Devaux Cave**

Joseph Devaux was an experienced and bold mountaineer. Exploring the range of the Pyrenees from the Pic with the help of a long equatorial scope, one day he observed a series of large funnels in the snow at the Cirque de Gavarnie, near l'Epaule du Marboré, which he explained by the presence of a fissure. He set off in mid-July 1928, discovered and explored a cave difficult of access and named it "cave of the sisters of the cascade", but which now bears his name. Devaux discovered a gallery whose walls were covered with a very thick layer of ice from which hung very large ice crystals. For him, the exceptionally large size of the crystals was due to the stability of the temperature in this gallery. He left minimal-maximal thermometers and returned several times, believing that methodical geological observations would provide remarkable discoveries.

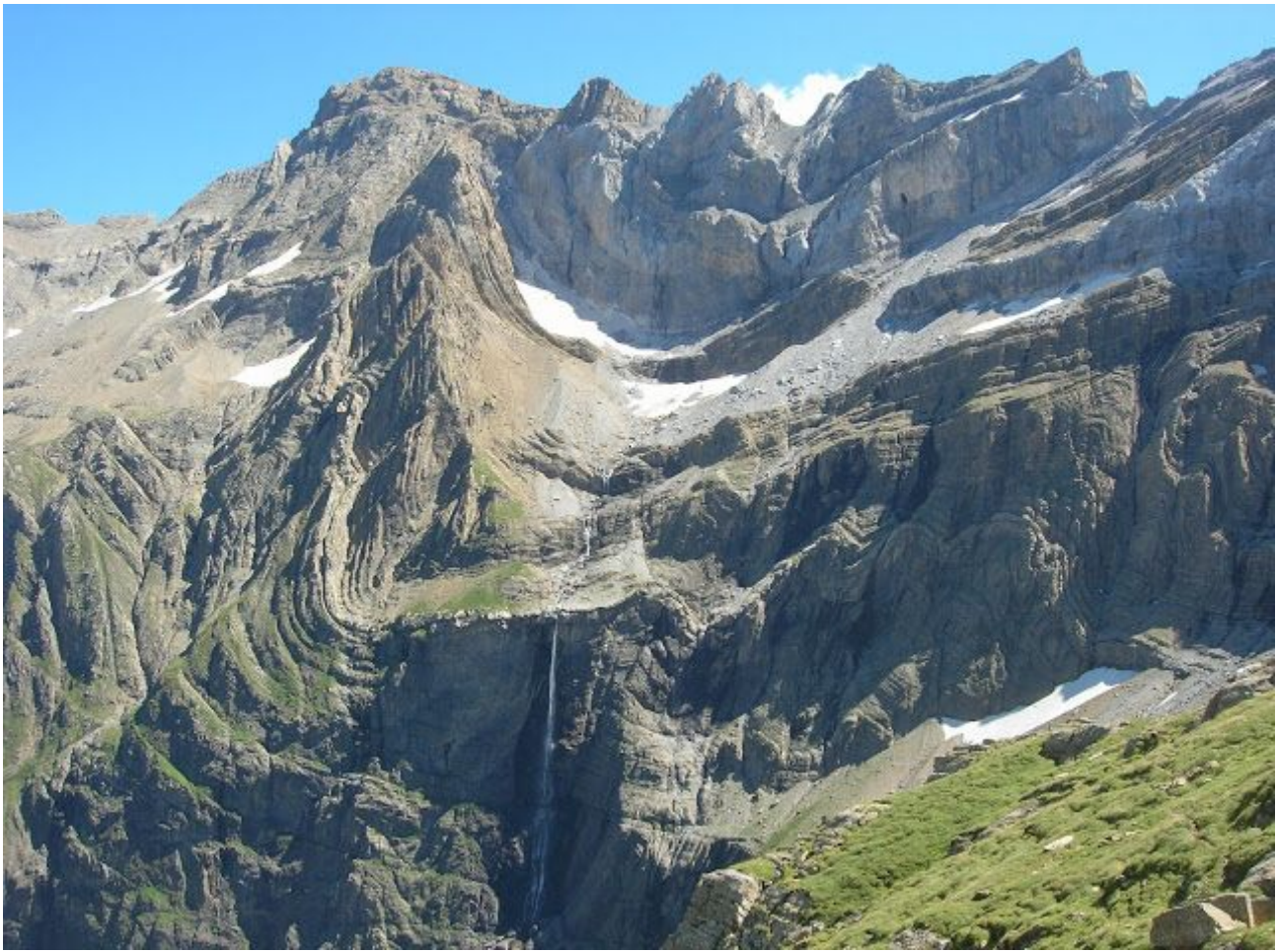

*Cirque de Gavarnie : the Devaux cave is below the rightmost of the three sharp peaks, at the level of the top glacier.*

Joseph Devaux died at the age of 34, when he had still not realized his full potential and was filled with scientific projects. He would certainly have had the means to fulfill them by replacing Camille Dauzère as director of the Pic in 1937. But rather than asking ourselves about a promising career prematurely interrupted, let us continue Joseph Devaux's personal journey since that summer of 1920. An unfailing desire to live at the Pic du Midi and to do science there, intense curiosity, publications supported by the Academy of Sciences (see below), the recognition of his value by men such as Charles Fabry, Charles Maurain and Jean-Baptiste Charcot, human qualities that made possible, and even harmonious and fruitful, life at the summit in difficult conditions. That was Joseph Devaux.

**Acknowledgments**. This is the translation of a paper published in *La Météorologie*, 1997, 8e série, n°18, 49-55. We thank Elise Devaux-Morin, sister of Joseph Devaux, for her help and for providing documents.

**References**

Davoust E. et J. Damiens, 1995, Histoire de l'observatoire du pic du Midi. *La Météorologie*, 8° série, numéro spécial, 22-27.

## Appendix : Publications of Joseph Devaux

Devaux J., 1927, Sur l'existence et la localisation de courants d'air verticaux ascendants aux environs du sommet du pic du Midi ; leur utilisation par les vautours. *Comptes-rendus de l'Académie des Sciences*, 184, 295-297.

Devaux J., 1927, Variation diurne de la température d'un champ de neige, unpublished, 17p.

Devaux J., 1927, Sur la mesure de la densité des champs de neige et des glaciers. *Comptes-rendus de l'Académie des Sciences*, 185, 1147-1149.

Devaux J., 1927, Sur la formation des glaciers par fusion diurne et regel nocturne des névés. *Comptes-rendus de l'Académie des Sciences*, 185, 1602-1604.

Devaux J., 1928, Sur le « rayon vert ». *L'Astronomie*, 42, 384-389.

Devaux J., 1929, Nouvelle grotte marboréenne. *La Nature*, 102.

Devaux J., 1929, Mesure du facteur d'absorption de la surface de quelques glaciers pyrénéens pour les radiations solaires. *Comptes-rendus de l'Académie des Sciences*, 188, 928-930.

Devaux J., 1929, Étude actinométrique de la pénétration du flux énergétique solaire à l'intérieur de quelques glaciers pyrénéens. *Comptes-rendus de l'Académie des Sciences*, 188, 1054—1055.

Devaux J., 1930, Étude photométrique de la pénétration des rayons solaires à l'intérieur des glaciers pyrénéens. *Comptes-rendus de l'Académie des Sciences.* 191. 1358—1360.

Devaux J ., 1931, Étude du rayonnement infrarouge émis par l'atmosphère terrestre. *Comptes-rendus de l'Académie des Sciences*, 193, 1207-1209.

Link F., J. Devaux, 1931, Étude photométrique et actinométrique de la Lune pendant l'éclipse du 26 septembre 1931. *Comptes-rendus de l'Académie des Sciences*, 193, 998—1000.

Maurain C., J. Devaux, 1932, Étude sur la conductibilité électrique et les noyaux de condensation atmosphérique au cours d'un voyage au Groenland. *Comptes-rendus de l'Académie des Sciences*, 195, 837—840.

Devaux J., 1933, Étude des glaces de terre et de mer au Scoresbysund (croisière 1932 du Pourquoi—Pas ?). *Annales hydrographiques*.

Devaux J., 1933, L'économie radiothermique des champs de neige et des glaciers. *Annales de physique*, volume 10, N° 20, 5-67.

Devaux J., 1933, L'économie radiothermique du globe terrestre. *66e congrès de sociétés savantes*, 412-416.

Devaux J., 1934, Étude du spectre infrarouge lointain du soleil. *Comptes-rendus de l'Académie des Sciences*, 198, 1595-1596.

Devaux J., 1935, Sur la température de l'ozone atmosphérique. *Comptes-rendus de l'Académie des Sciences*, 201, 1500.

Devaux J., 1936, Sur un nouveau galvanomètre utilisé dans les montages à résonance pour l'infrarouge. Galvanomètre autostabilisé (sans dérive) à double cadre mobile. *Le journal de physique et le radium*, série 7, tome 7, 3, 146-148.